# Energy harvesting from coupled bending-twisting oscillations in carbon-fibre reinforced polymer laminates

Mengying Xie[1], Yan Zhang[1], Marcin J. Kraśny[1], Andrew Rhead[1], Chris Bowen[1] and Mustafa Arafa[2]

[1] Mechanical Engineering Department, University of Bath, Bath A2 7AY, UK

[2] Mechanical Engineering Department, American University in Cairo, Egypt

**Highlight**

- First demonstration of laminated beam with bend-twist coupling for energy harvesting
- Unique approach to tune the frequency response of the resonant harvesting system
- Novel laminate structure and model tuned to provide a broader-band energy harvester

**Abstract**

The energy harvesting capability of resonant harvesting structures, such as piezoelectric cantilever beams, can be improved by utilizing coupled oscillations that generate favourable strain mode distributions. In this work, we present the first demonstration of the use of a laminated carbon fibre reinforced polymer to create cantilever beams that undergo coupled bending-twisting oscillations for energy harvesting applications. Piezoelectric layers that operate in bending and shear mode are attached to the bend-twist coupled beam surface at locations of maximum bending and torsional strains in the first mode of vibration to fully exploit the strain distribution along the beam. Modelling of this new bend-twist harvesting system is presented, which compares favourably with experimental results. It is demonstrated that the variety of bend and torsional modes of the harvesters can be utilized to create a harvester that operates over a wider range of frequencies and such multi-modal device architectures provides a unique approach to tune the frequency response of resonant harvesting systems.

**Nomenclature**

| | |
|---|---|
| $C$ | Damping matrix |
| $F$ | Force |
| $I$ | Effective mass moment of inertia |
| $k$ | Stiffness |
| $k_T$ | Torsional stiffness |
| $[K]$ | Stiffness matrix |
| $l$ | Arm length denoting bend-twist coupling |
| $m$ | Effective mass |
| $[M]$ | Mass matrix |
| $t$ | Time |
| $T$ | Kinetic energy |
| $V$ | Potential energy |
| $x$ | Transverse displacement of tip cross-section |
| $y$ | Base motion |
| $Y$ | Amplitude of base motion |
| $\alpha$ | Damping coefficient for mass matrix |
| $\beta$ | Damping coefficient for stiffness matrix |
| $\delta$ | Nodal degrees of freedom |
| $\phi$ | Cross-section rotation due to torsional loads |
| $\theta$ | Cross-section rotation under no torsional loads |
| $\omega$ | Angular frequency |

## 1. Introduction

The conversion of mechanical vibrations into useful electrical energy has been a subject of intensive research due to its application in self-powered sensors and wireless systems. A common architecture for vibration-based energy harvesting devices is a base-excited elastic structure, such as a mass-spring system or a cantilever beam, which is typically used in conjunction with electromagnetic and piezoelectric devices. One of the challenges in designing continuous systems, such as beams and plates, for piezoelectric energy harvesting applications lies in the placement of the energy generating materials, which are usually piezoelectric sheets or layers that are bonded or embedded within a host structure. These structures often have unique strain distributions in their vibration modes, and the highly strained parts of the structure that are most effective for energy harvesting occur only in localized regions. The placement of the energy generation materials at such discrete locations while leaving other parts of the structure uncovered causes the structure to be partially utilized for energy harvesting, which lowers the power density. This is the case, for example, for a cantilever beam undergoing flexural vibration at its first bending mode, where maximum strain occurs near the root. To overcome this difficulty, systems that operate across multiple vibration modes have been proposed [1, 2] in order to utilize two or more vibration mode shapes for energy harvesting. The exploitation of more than one vibration mode also enables the device to harvest vibrations over a wider frequency range since in many cases the ambient vibrations to be harvested often span a range of frequencies and amplitudes.

Vibration modes involving mixed deformations, in particular bending and twisting motions, are particularly appealing in this respect since these motions can be controlled by design of the harvesting structure. The study of coupled bend-twist oscillations has been of interest to the aeronautical engineering community for decades owing to its application in the vibration analysis of aircraft wings and rotating turbine blades. Recent interest in morphing structures has spurred research in the use of piezoelectric actuators to achieve a controlled bend-twist deformation [3-5]. The use of bend-twist oscillations in energy harvesting applications is relatively recent. In this context, reference is made to the work of Abdelkefi *et al*. [6] who designed a unimorph cantilever beam undergoing bending–torsion vibrations consisting of a single piezoelectric layer and two asymmetric tip masses, thereby generating a twisting moment from a base excitation. This structure was tuned to provide a broader band energy harvester by adjusting the first two global natural frequencies to be relatively close to each other. Reference is also made to the work of Gao *et al*. [7] in which torsional vibration at the second mode of a cantilever beam with an eccentric proof mass was employed for energy harvesting using a lead zirconate titanate (PZT) material. The main advantages of this design approach were the small displacement amplitudes and low natural frequency. Shan *et al*. [8] employed vortex-induced vibrations to design a piezoelectric energy harvester with bending-torsion vibration.

An effective way of designing structures with inherent bend-twist coupling behavior is the use of composite laminates. By tailoring the laminate lay-up these anisotropic structures can be deliberately designed to exhibit interactions between extension, shear, bending and twisting [9], which are not present in conventional isotropic materials. In this work, we shall employ for the purpose of energy harvesting, a laminated carbon fibre reinforced polymer (CFRP) composite cantilever beam that has a laminate lay-up that is selected to achieve coupled bending and twisting deformations. Specifically, this includes the use of a laminate with an unbalanced stacking sequence in which not all plies with a positive rotation in the stacking sequence have a counterpart ply with an equal and negative rotation. Such an unbalanced structure introduces extension-shear coupling and when the laminate (or a single ply with fibres misaligned with the load direction) is subject to a unaxial tensile or compressive load it will attempt to shear. The sign of the shear is dependent on whether the load is tensile or compressive. Under a bending deformation, loading changes from being compressive on one side of the neutral axis to tensile on the other and this leads to plies on opposite surfaces attempting to shear in opposite directions, thereby resulting in twisting of the laminate. This approach is particularly attractive since it removes the need for more complex design configurations; such as the use of an eccentric proof mass or asymmetric tip masses. The ability to tailor the bend-twist coupled laminate architecture also provides scope for a wide design space to tailor the cantilever response to the vibration spectrum being harvested. To utilize larger portions of the beam for power generation, we employ two different types of PZT materials that are attached to the beam's surface. One PZT patch (MFC M8557-P1) responds to uniaxial straining (extension-mode) since the active piezoelectric material is aligned along its length and is placed close to the beam's root for effective harvesting of the bending mode. The other material (MFC M8557-F1) operates in the shear-mode, and in this case the piezoelectric is aligned at 45° to its length, and is placed along the beam's mid span where torsional shear strains are higher. This effort is expected to eventually yield energy harvesting devices with greater power densities, and greater bandwidths which are increasingly in high demand for small sensors and other applications where miniaturization is of the essence.

The remainder of this paper is organized into four sections. Section 2 details the fabrication of the laminated composite beams. Section 3 presents a mechanical model of the device and provides an analysis of its frequency response. Experimental work is carried out in Section 4. Finally, Section 5 is dedicated to conclusions and recommendations.

2. **Fabrication of Composite Bend-Twist Laminate**

The cantilever beam used in this work was made of unidirectional CFRP plies (Cytec MTM 49-3/ T800HB-140-36%). Plies were laid down by hand in the order of the stacking sequence with debulking under vacuum applied after every 4$^{th}$ ply to remove air pockets. Once all plies were laid the laminate was cured in an autoclave at 125°C with a pressure of 7 bar for 2 hours using a vacuum bag moulding process. A rigid aluminium plate was placed on top of the laminate and a flexible 3 mm thick silicone rubber sheet were used at bottom and top to form smooth outer surfaces. Figure 1 shows a schematic of the cantilever beam, together with its dimensions and the position of the piezoelectric patches. The cantilever consists of 10 symmetric plies with the layup of [0, 30, 0, 45, -45]$_s$, where 0° ply is along the span of the beam. This laminate system was chosen since (i) the symmetric nature of the laminate ensures that it will remain flat and warp free after high temperature curing, and (ii) to achieve bend-twist coupling the laminate should have an unbalanced lay-up, as described above. The width of the cantilever beam was 100 mm and the thickness was

1.45 mm after curing. The beam had a free length of 499 mm with an additional 60 mm for clamping between two aluminium plates.

The PZT materials used in this work are Macro Fibre Composites (MFC) which consist of PZT fibers embedded in an epoxy matrix with interdigitated electrodes. One PZT patch (Smart Material, MFC M8557-P1) responds to uniaxial straining, with piezoelectric fibres aligned along its length, was placed close to the beam's root (at a distance of 24.5 mm from the beam's fixation, as shown in Figure 1). The second patch (Smart Material, MFC M8557-F1) operates in the shear-mode where the piezoelectric fibres in the MFC are at 45° to its length axis. This MFC was placed along the beam's mid span (at a distance of 68.2 mm from the root patch, as shown in Figure 1) where torsional shear strains are higher.

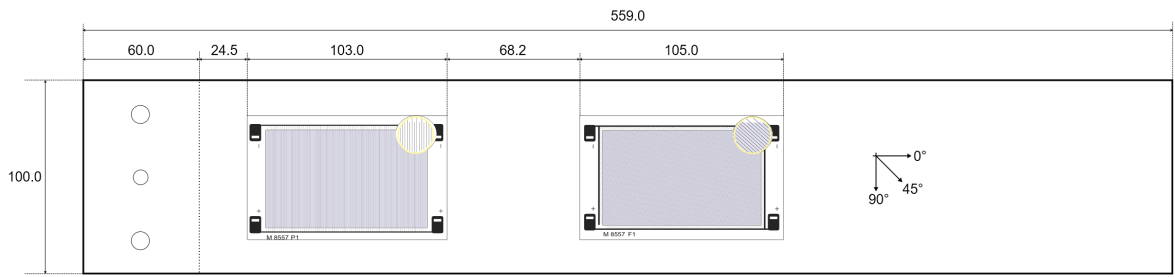

Figure 1. Schematic of the cantilever beam and position of piezo-patches. The cantilever was clamped at left hand side. Near the root the piezoelectric element was a MFC M8557-P1 (uniaxial mode), and an MFC M8557-F1 (shear mode) at the mid-span.

The stiffness of the cantilever beam was experimentally determined by statically loading a clamped-free beam at centre of its tip on an Instron 3365 tensile testing machine equipped with a 100 N load cell, while measuring the tip deflection, which is labelled as $x$ in Fig. 4. Figure 2 shows the resulting force-deflection behaviour, which clearly indicates a linear relationship up to 50 mm of deflection, which corresponds to almost 3 N of applied force.

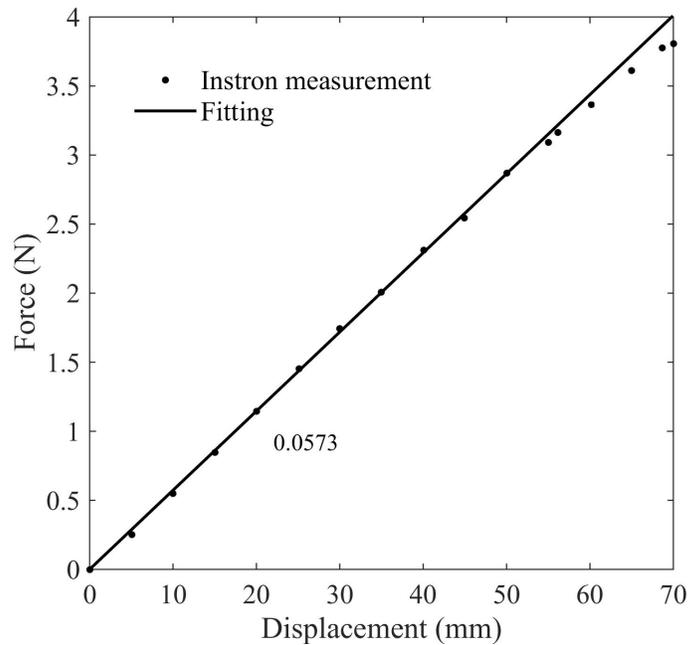

Figure 2. Force-deflection curve of cantilever indicating a linear response for a deflection up to 50mm (~3N).

Due to the bend-twist coupling, the tip of the beam developed an angle of twist $\theta$ that increased with the tip displacement, as shown in Figure 3(a). Figure 3(b) demonstrates that for deflections less than 50 mm, the twisting angle increased linearly with displacement, which correlates with the force-displacement observations. It can be noticed when the displacement was zero, the tip of the beam had small twist, which is a result of the residual stresses in the laminate due to unbalanced layup. However, both force and angle had a nonlinear behaviour at displacements above 50 mm, and beyond 50 mm twisting became more obvious than bending.

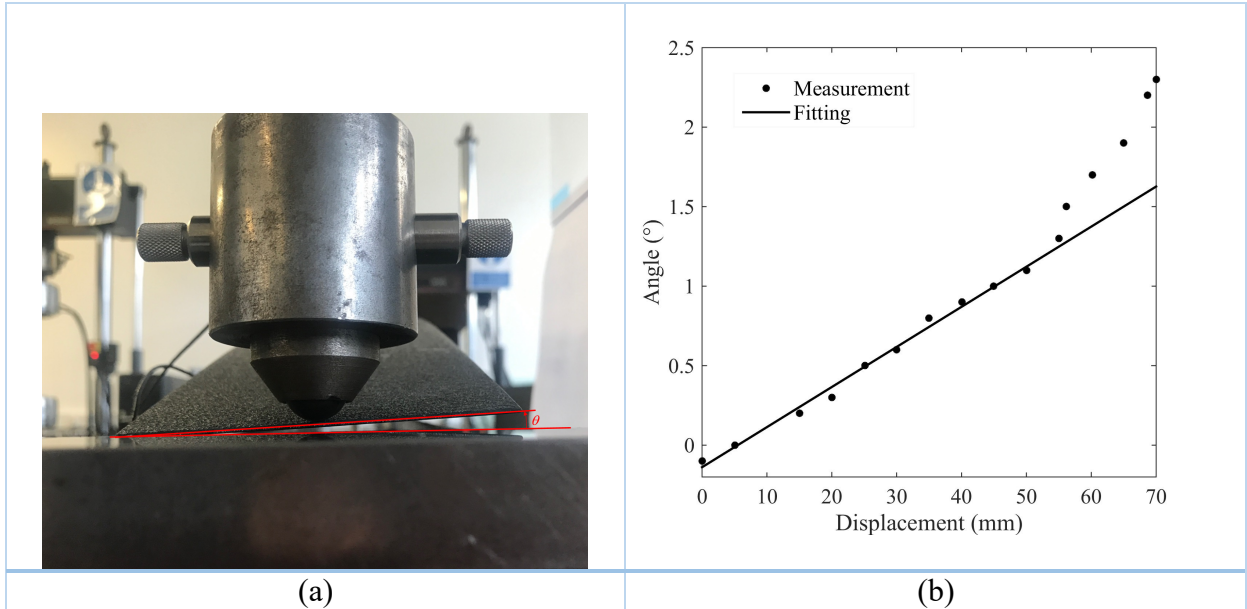

| (a) | (b) |

Figure 3. Bend-twist coupling of the cantilever beam. (a) Image of cantilever undergoing bend-twist coupling, $\theta$ is the twisting displacement (b) Relationship between twisting angle and deflection (linear below 50mm, slope=0.0252°/mm).

## 3. Mechanical Model

Figure 4 shows a schematic of the cantilever beam with the two PZT patches undergoing bending and twisting deflections. While structures undergoing coupled bend-twist deformations are essentially two dimensional, the aim of this work is to develop a simplified lumped-parameter model to enable the analysis of the system as a two degree of freedom (2-DOF) model. In this way, the transverse displacement $x$ of the centroid of the cross-section at the tip, and the angular displacement $\theta$ of the tip section are taken as representative degrees of freedom that designate bending and twisting deformations of the cantilever beam, respectively.

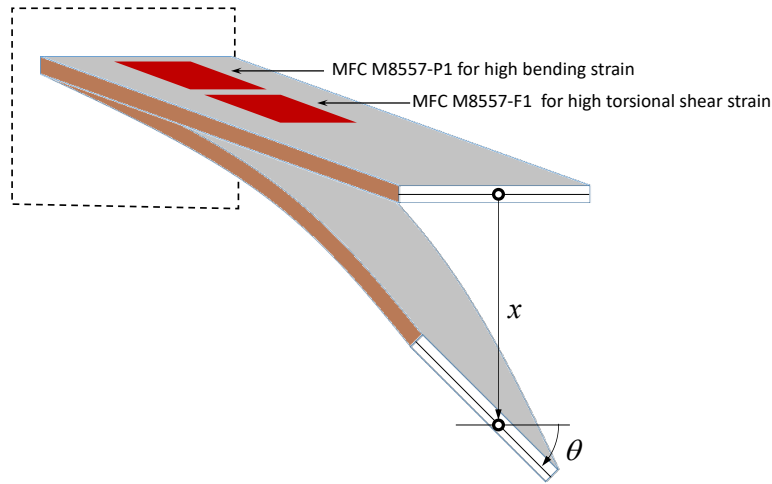

Figure 4. Schematic of a deformed of cantilever beam of a rectangular cross-section undergoing a coupled bend-twist deformation pattern featured by a transverse displacement x and an angle of twist θ of the tip cross section.

The laminated composite beam is modelled as an equivalent rigid body of mass *m* and mass moment of inertia *I*, as shown in Figure 5. The body is attached to a base by a rigid massless arm of length *l*. A linear spring of stiffness *k* connects the mass to the base, whereas a torsional spring of stiffness $k_T$ connects the mass to the link. This simplified model provides some physical insight into the problem and allows a study of the effect of the individual design parameters on the system response. The beam is modeled as a 2-DOF mass that undergoes coupled bend-twist oscillations. The bending displacement is denoted by *x* and resembles the transverse displacement of the beam tip. The twist displacement is denoted by *θ* and represents the angle of twist of the beam tip cross-section under no external torsional loads. Both degrees of freedom are piezoelectrically coupled and the electrical outputs are connected to resistive loads.

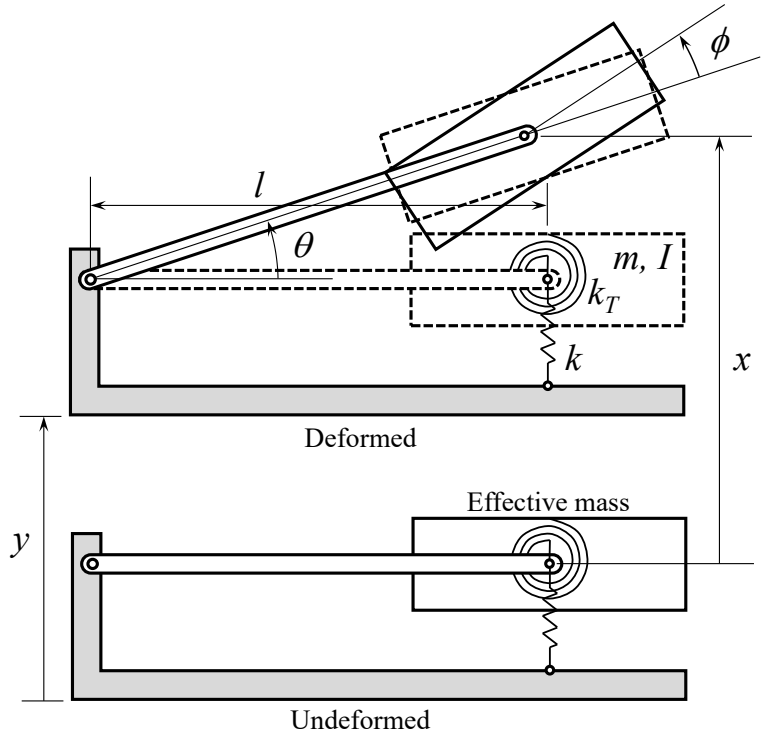

Figure 5. Lumped-parameter model of the bend-twist coupled system.

The total kinetic energy of the system is given by:

$$T = \frac{1}{m}m\dot{x}^2 + \frac{1}{2}I(\dot{\theta} + \dot{\phi})^2 \tag{1}$$

where *m* and *I* are the effective mass and mass moment of inertia, respectively, and the angle *ϕ* represents the cross-section rotation due to torsional loads. These will typically arise as a result of backward coupling twisting moments generated by the torsional MFC M8557-F1 patches. From geometry, it can be observed that:

$$(x - y) = l\theta \tag{2}$$

where $l$ denotes the arm length that designates the inherent bend-twist coupling. Equation (1) can hence be expressed as:

$$T = \frac{1}{2}m\dot{x}^2 + \frac{1}{2}I\left(\frac{\dot{x}}{l} - \frac{\dot{y}}{l} + \dot{\theta}\right)^2 \tag{3}$$

The total strain energy of the system is given by:

$$V = \frac{1}{2}k(x-y)^2 + \frac{1}{2}k_T\phi^2 \tag{4}$$

where $k_T$ is the torsional stiffness of the beam. Using Lagrange's equations, the equations of motion of the system are obtained as:

$$\begin{bmatrix} m + I/l^2 & I/l \\ I/l & I \end{bmatrix} \begin{Bmatrix} \ddot{x} \\ \ddot{\phi} \end{Bmatrix} + \begin{bmatrix} k & 0 \\ 0 & k_T \end{bmatrix} \begin{Bmatrix} x \\ \phi \end{Bmatrix} = \begin{Bmatrix} ky + I\ddot{y}/l^2 \\ I\dot{y}/l \end{Bmatrix} \tag{5}$$

Equation (5) can be written in a general form as:

$$[M]\{\ddot{\delta}\} + [K]\{\delta\} = \{F\} \tag{6}$$

Damping is arbitrarily introduced through a proportional damping matrix in the form:

$$[C] = \alpha[M] + \beta[K] \tag{7}$$

where $\alpha$ and $\beta$ are damping coefficients for the mass and stiffness matrices, respectively. Equation (7) describes the mechanical behaviour of the beam and can be used to obtain the steady-state response under a sinusoidal base excitation of the form $y(t) = Y \sin\omega t$. The system parameters are obtained from experimental observations and, whenever possible, from equivalent lumped-parameter models. Table 1 lists the parameters used in this study. Values of the damping coefficients $\alpha$ and $\beta$ have been chosen to provide quality factors match those observed experimentally.

Table 1. System parameters

| Parameter | Value |
|---|---|
| $m$ | 35.28 g |
| $I$ | $5 \times 10^{-5}$ kgm² |
| $l$ | 2.3 m |
| $k$ | 57 N/m |
| $k_T$ | 4 Nm/rad |
| $\alpha$ | 3 |
| $\beta$ | 0 |

## 4. Experimental Work

The MFC device is a flexible lead zirconate titanate (PZT) based sheet of aligned rectangular piezoelectric fibres sandwiched between epoxy layers and interdigited electrodes with a manufacturer's specified capacitance of 12.84 nF and 13.26 nF for M8557-P1 and M8557-F1, respectively [10]. The M8557-P1 patch is designed to be affected by expanding motion (extension-mode), while M8557-F1 with fibres rotated at 45° to the root was used to respond to the beam's twisting motion (shear-mode).

The cantilever beam was mounted vertically on to an electrodynamic shaker (LDS V455) by two aluminium plates. Figure 6a shows a schematic of the experimental setup and Figure 6b shows an image of the system, along with the cantilever beam. An electrodynamic shaker was controlled by an amplified (LDS PA 1000) signal generated by NI USB-6211 (National Instruments) with a dedicated LabView script. The shaker system was calibrated using a Lab View routine to provide constant acceleration at the desired range of frequencies. An LDS 455 shaker was calibrated by exciting the shaker with a sine wave of controlled amplitude and frequency running through the LDS PA 1000 amplifier. A matrix of voltage amplitudes and frequencies was sent to the shaker, and the response was measured using a Polytech laser differential Vibrometer which then measured the velocity using the PSV-400-M4 scanning head, OFV-5000 controller and VD-09 velocity decoder. The peak acceleration was calculated from the velocity data and saved in the calibration file using a LabView routine running on a computer. When a specific acceleration is required at a desired frequency, another routine interpolates between the nearest voltages to deliver the needed acceleration magnitude [11].

The electrical output from the MFCs was connected to electrical load, set by resistance decade (JJ JUNIOR Resistance Decade from Educational Measurements Ltd). Finally, the signal was gathered by oscilloscope (Agilent Infiniium 54835A with 10MΩ/10pF TEK P6137 oscilloscope probe).

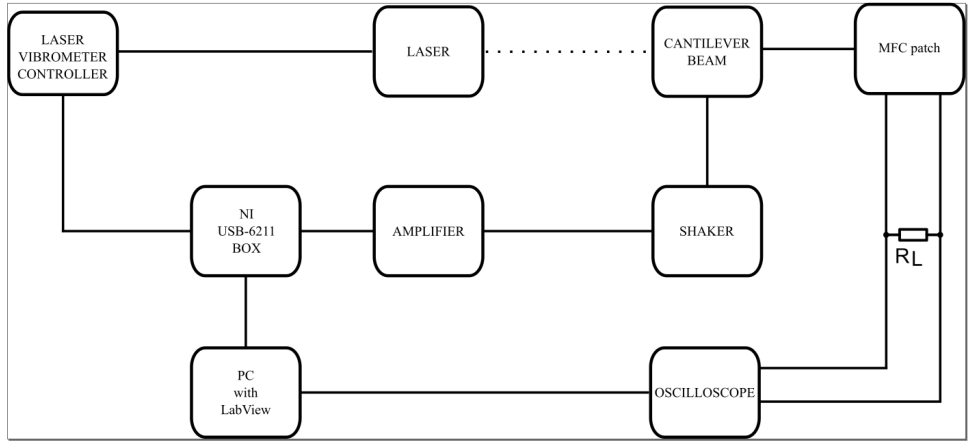

(a)

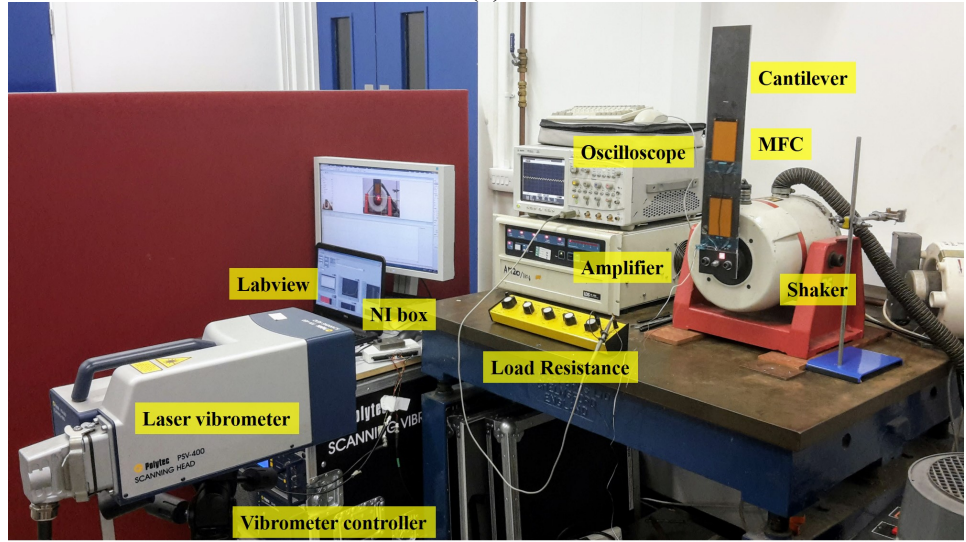

(b)

Figure 6. Experimental setup. (a) Schematic of experiment setup. R$_L$ is the load resistance (b) Experiment setup showing bend-twist cantilever beam that is mounted vertically on to the shaker.

The power output from the MFC patches was determined from the equation,

$$P_{OUT} = U_{RMS}^2/R_L \tag{8}$$

where $U_{RMS}$ is the root mean square of the voltage output across the load resistor and $R_L$ is electrical resistance of the load resistor; in this case $R_L$=10 kΩ.

The power output generated by the MFC piezoelectric materials is sensitive to the matching impedance, which is dependent on the excitation frequency. The optimum load impedance fulfils the condition delivered from equation for a capacitive reactance,

$$\omega R_L C_p = 1 \tag{9}$$

which can be transformed to form:

$$R_L = \frac{1}{2\pi C_p f_t} \tag{10}$$

where $C_p$ is a capacitance of the piezoelectric device, $\omega = 2\pi f_t$ is the angular frequency, and $f_t$ is the desired testing frequency from the frequency range of interest. For the sweep function in the frequency range of interest between 6-45 Hz, the optimum load impedance varied from 2 MΩ to 267 kΩ. In order to minimize the influence of the matching impedance, the value of 10 kΩ was used throughout the harvesting experiments.

## 5. Results and discussion

For energy harvesting characterization, the electrodynamic shaker was excited at a constant 1g (9.8 m/s²) and the vibration frequency was swept from 5 to 52 Hz to cover a range of vibration modes. Figure 7 shows the measured power output from both the extension-mode (M8557-P1) and shear-mode (M8557-F1) MFC patches. The extension-mode MFC attached on the root of the cantilever generated power at a lower frequency of 6.5 Hz only, which reached a peak of 1.15 mW. However, the shear-mode MFC at the mid span had a small power peak at 6.5 Hz and two higher peak powers at 37 and 45 Hz (0.7 mW and 0.4 mW, respectively). Since the MFC on the root harvested the bending and the MFC in the mid-span harvested the twisting of the cantilever beam, the bending mode and twisting mode can be characterized experimentally and Figure 7 indicates that the bending mode is 6.5 Hz and the twisting mode is 37 and/or 45 Hz. The small peak in power at 6.5 Hz from the shear-mode MFC is thought to be due to the bending of the beam.

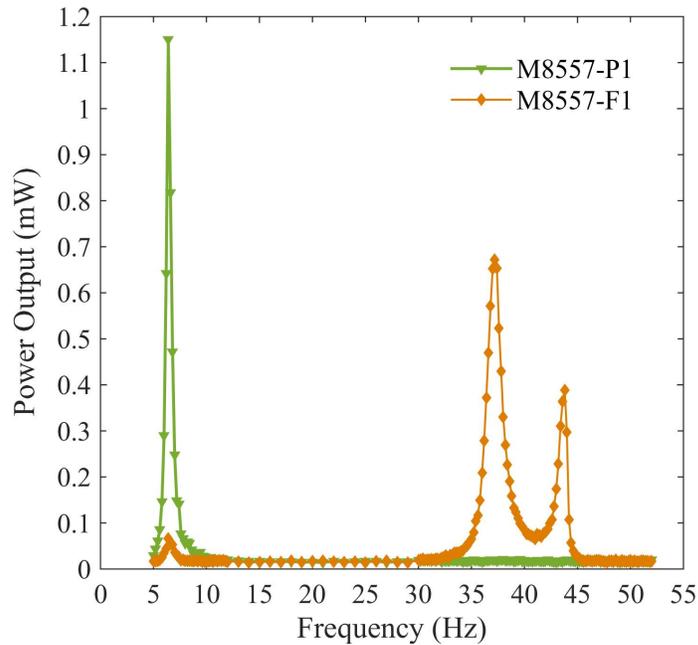

Figure 7. Power output from MFC patches at excitation level of 1g in frequency range of 5-52 Hz. Green(▼) output from M8557-P1 (extension-mode), red (♦) output from M8557-F1 (shear-mode).

Figure 8 shows the prediction of the 2-DOF model as the steady-state response of the tip displacement relative to the base motion ($x$-$y$) and angle of twist ($\phi$) for the range of frequencies studied experimentally. It can be observed that the 2-DOF model predicts resonance frequencies at 6.4 Hz and 45 Hz, which compare very favorably with the experimental values in Figure 7. The first mode is dominated by transverse deformation, which explains the high electrical output of the extension-mode PZT patch, and the other mode is dominated by twist, which is expected to yield an output from the shear-mode PZT patch. The third resonance obtained experimentally at 37 Hz is a mixed mode that combines transverse deformation and twist and could not be captured by the present 2-DOF model, but indicated the potential of the harvesting system to operate at multiple modes.

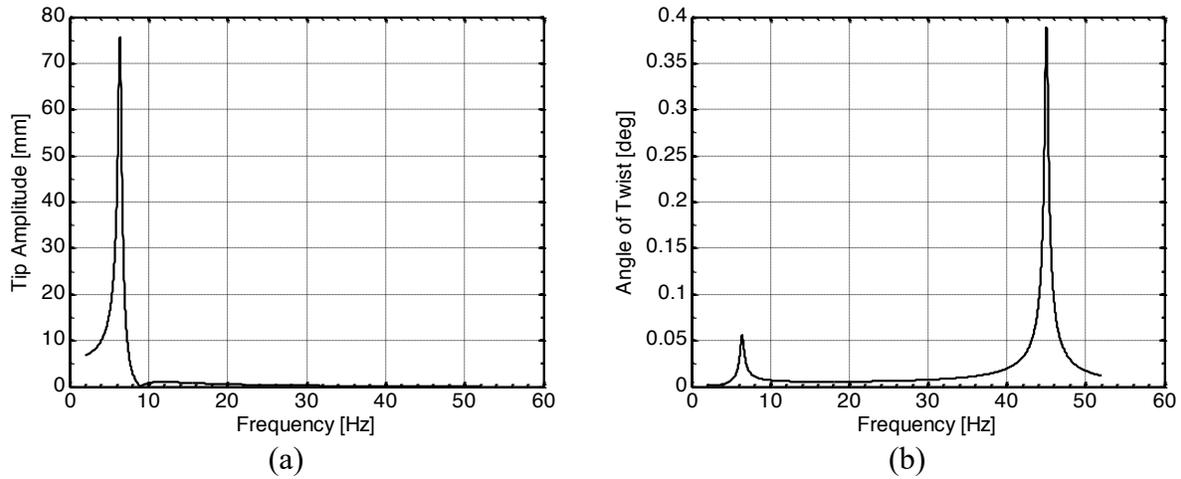

Figure 8. Simulated frequency response of the 2-DOF model at 1g excitation level. (a) Variation of the tip displacement amplitude with respect to the base motion ($x$-$y$); (b) angle of twist $\phi$.

Figure 9 schematically illustrates the vibration mode shapes observed experimentally by viewing the beam through a strobe light. It can be observed that the deformation pattern in the first mode (7 Hz) is predominantly bending. The ensuing uniaxial strains, which are greatest at the root, leads to a greater electrical output from the extension-mode PZT patch, the lower patch in Figure 6b. The shear-mode PZT patch (upper patch in Figure 6b) is not subjected to significant torsion and hence provides minimal output, as shown in Figure 7. The second mode (37 Hz) was observed to contain primarily twisting motions, superimposed on a bending deformation pattern that is somewhat close to the second bending mode in a homogeneous cantilever beam. In this mode, bending deformations at the root are small, hence output from the lower patch is insignificant, whereas torsional deformations provide a higher output from the upper patch. The third mode (45 Hz) is dominated by twisting, which provides favourable drive conditions for the shear-mode patch and no output from the extension-mode one, which agrees with the output shown in Figure 7.

While no proof mass was used in this study they are often employed in cantilever-type energy harvesters to maximize the power output without significantly increasing the overall device volume. As the unbalanced CFRP beams presented here have inherent characteristics that induce bend and twist, adding a proof mass can further enhance bending and twisting effects. It is envisaged that the introduction of point masses in the form of asymmetric masses will induce even more twisting, which will provide favourable output from the shear-mode PZT patch.

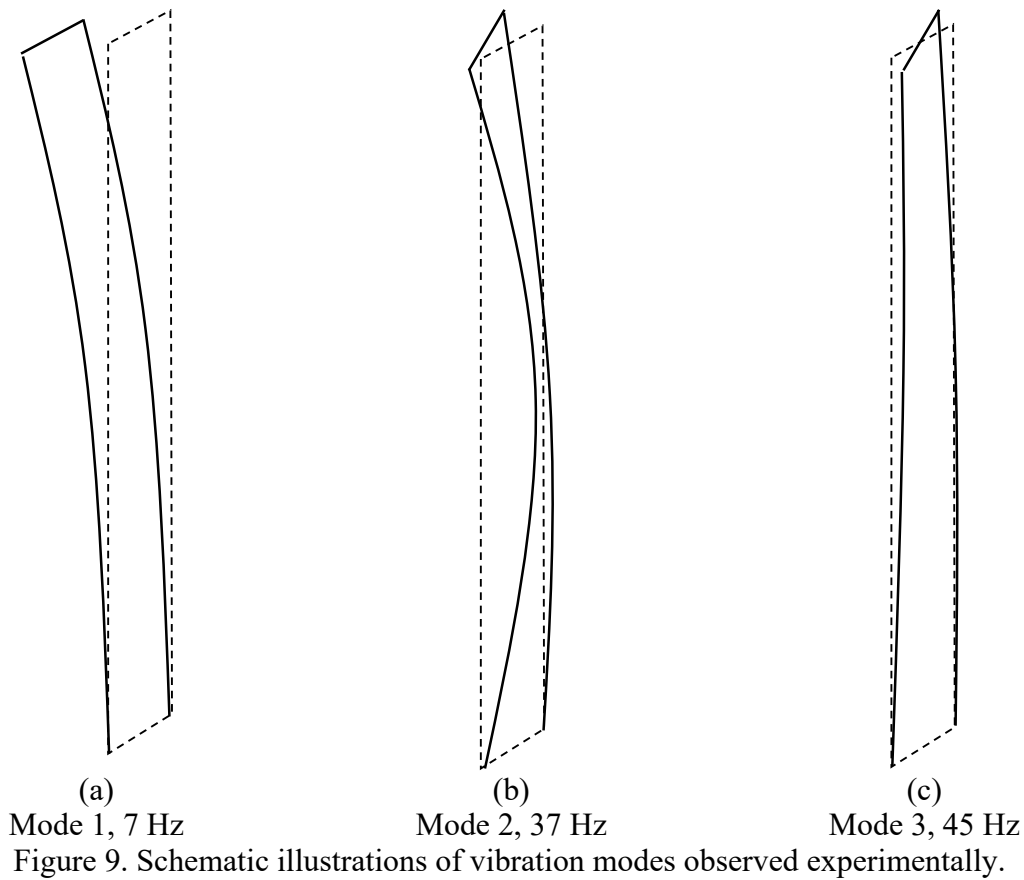

| (a) | (b) | (c) |
| Mode 1, 7 Hz | Mode 2, 37 Hz | Mode 3, 45 Hz |

Figure 9. Schematic illustrations of vibration modes observed experimentally.

Accordingly, the present beam possesses multiple, narrow-band harvestable vibration modes exploiting both bending and torsional deformations. Through design optimization, the spectral spacing between these modes can be minimized to offer a multi-modal device that operates across a defined frequency band, which can be effective for applications where the excitation frequency is known to vary within certain limits.

**Conclusions**

This work has presented the first demonstration of the use of a laminated carbon fibre reinforced polymer to create cantilever beams with coupled bending-twisting oscillations for multi-mode energy harvesting applications. Modelling of this new bend-twist harvesting system in presented, which compares favourably with the experimental results. It is demonstrated that the variety of bend and torsional modes of the harvesters can be utilized to create a harvester that operates in a wider range of frequencies and such device architectures provides a unique approach in which to tune the frequency response of the resonant harvesting system. The advantages of the proposed design is that the structure can be tuned to provide a broader-band energy harvester by adjusting natural frequencies. In addition, torsional modes typically exhibit small displacement amplitudes compared to bending modes; this provides scope for smaller volume device architectures. By exploiting at laminate architecture which inherently exhibits bend-twist coupling the complexity of the system is reduced since there is no need to use an eccentric proof mass or asymmetric tip masses to achieve torsion and the numerous potential lay-ups that exhibit bend-twist coupling provides a large design space to optimise the frequency response of the cantilever. This approach therefore provides new opportunities for the development of broader-band energy harvesting materials and systems.

**Acknowledgement**

This work was supported by both the European Commission's Marie Skłodowska-Curie Actions (MSCA), through the Marie Skłodowska-Curie Individual Fellowships (IF-EF) (H2020-MSCA-IF-2015- EF-703950-HEAPPs) under Horizon 2020 and the European Research Council under the European Union's Seventh Framework Programme (FP/2007–2013)/ERC Grant Agreement no. 320963 on Novel Energy Materials, Engineering Science and Integrated Systems (NEMESIS).